\documentclass{PoS}

\usepackage{amsmath}
\usepackage{amssymb}
\usepackage{graphicx}
\usepackage{xspace}
\usepackage{url}
\usepackage[utf8]{inputenc}
\usepackage{examplep}

\newcommand{\myurl}[1]{{\footnotesize \href{#1}{\PVerb{#1}}}}
\newcommand{\arXiv}[1]{\href{https://arxiv.org/abs/#1}{arXiv:#1}}
\usepackage{txfonts} 

\usepackage{booktabs}

\newcommand{\ab}{\;\mathrm{ab}}
\newcommand{\fb}{\;\mathrm{fb}}

\newcommand{\MeV}{\;\mathrm{MeV}}

\newcommand{\Hff}{H\!f\!f}

\newcommand{\Dslash}{\hspace{0.2em}/\hspace{-0.6em}D}

\usepackage{xcolor}
\definecolor{light-gray}{gray}{0.8}

\title{``Theory Vision'' talk at LHCP2018}

\ShortTitle{Theory Vision}

\newcommand{\OXaff}{Rudolf Peierls Centre for Theoretical Physics,
  Clarendon Laboratory, Oxford OX1 3PU, UK}
\newcommand{\ASCaff}{All Souls College, Oxford OX1 4AL, UK}
\newcommand{\CERNaff}{CERN,
    Theoretical Physics Department, CH-1211 Geneva 23, Switzerland}

\author{Gavin P. Salam\thanks{On leave from CNRS, UMR 7589,
    LPTHE, F-75005, Paris, France.}\\
  \OXaff,\\
  \ASCaff,\\
  \CERNaff.
        \\
        E-mail: \email{gavin.salam@physics.ox.ac.uk}}

\usepackage{chngcntr}
\counterwithout{equation}{section}

\abstract{%
  Particle physics is sometimes described as going through a crisis,
  notably because of the continued lack of discovery of physics beyond
  the Standard Model, despite the LHC having operated at close to
  maximal energy for some years now.
  Here, I argue that we should not underestimate the significance of
  recent progress and future prospects in the Higgs sector of the
  Standard Model.
  This is especially the case for the Yukawa interactions and the
  structure of the Higgs potential, both of which are unlike any
  sector that has been established and stress-tested before in
  particle physics.
  Other topics that I touch on include the still substantial scope for
  increasing the reach of searches at LHC, the increasing role of
  precision in hadron-collider physics and the rich interplay that is
  developing between heavy-ion and proton-proton physics.
}

\FullConference{Sixth Annual Conference on Large Hadron Collider Physics (LHCP2018)\\
		4--9 June 2018\\
		Bologna, Italy}

\begin{document}

\section{Introduction}

A typical ``Vision'' talk often addresses the big, unanswered
questions, such as the nature of dark matter and dark energy,
questions about fine tuning and the origin of the matter-antimatter
asymmetry of the universe.
It is much less likely to discuss the standard model (SM) other than,
sometimes, to briefly state that the SM is now complete.


Searching for answers to big unanswered questions is vitally
important.
In doing so, however, we should not forget the importance of big
\emph{answerable} questions and the issue of how we go about answering
them.
That, for the most part, is what this talk is about.

The starting point is to examine the SM and ask what we genuinely know
about it, i.e.\ to what extent it is as complete as is often stated.
The SM, as a theoretical construction, is not just about its particle
content but, just as importantly, about a set of interactions.
Those interactions are summarised on a famous CERN T-shirt in the
following compact form:
\begin{align}
  \label{eq:1}
  \mathcal{L} = -\frac{1}{4} F_{\mu\nu} F^{\mu\nu}
  + i \bar \psi \Dslash \psi
  + \left| D_\mu \phi\right|^2
  + \psi_i y_{ij} \psi_j \phi + \text{h.c.}
  - V(\phi)\,.
\end{align}
That T-shirt comes with an accompanying card that says ``this
equation neatly sums up our current understanding of fundamental
particles and forces''.
What is the meaning of ``understanding'' in this context? Is it
knowledge? Is it an assumption?
Depending on the part of the Lagrangian that you examine, the answer
differs.

The pure gauge $F_{\mu\nu} F^{\mu\nu}$ and matter-gauge
$i \bar \psi \Dslash \psi$ terms are arguably knowledge.
The $i \bar \psi \Dslash \psi$ term encodes interactions such as
$ee\gamma$, $e\nu W$, $qqg$, $qqZ$ that have been established to high
accuracy in $e^+e^-$, DIS and hadron-collider experiments.
In terms of interactions, $F_{\mu\nu} F^{\mu\nu}$ term encodes in
particular triple-gauge interactions such as $ggg$ and $ZWW$ that have
similarly been probed experimentally at a range of past and present
colliders.
When one thinks of the immense success of the SM over the past 40
years, it is in most cases a success involving the gauge sector, a
sector whose history dates back, in some sense, to Maxwell's
equations.
If we measure yet another of the myriad individual contributions to
these structures, considering all possible flavour and gauge-boson
combinations, it is an important contribution to validating the
overall framework of the SM, but it is perhaps not surprising if we
are underwhelmed by the success, yet again, of a test of the SM.

The rest of Eq.~(\ref{eq:1}) deals with the Higgs sector of the
Standard Model, which was discussed extensively during LHCP (see for
the example the talks by Grefe, Ortona, Sanz and Vryonidou~\cite{Grefe,Ortona,Sanz,Vryonidou}).
I will spend quite some time discussing that in the next section.
That will be followed by a briefer discussion of searches and anomalies,
precision standard-model theory and heavy-ion collisions.

\section{The Higgs sector of the Standard Model}

Of the remaining terms in Eq.~(\ref{eq:1}), one of them,
$\left| D_\mu \phi\right|^2$, which again involves a gauge
interaction, but now with the Higgs field, can also be argued to be
knowledge, albeit with less precision than other gauge terms.
It encodes two interactions, $HWW$ and $HZZ$, both of which
contributed to the Higgs-boson discovery and continue to see extensive
study~\cite{Ortona}. 
In a sense, this part is a variation on the other well-established
gauge terms of the SM Lagrangian.

The last two terms of Eq.~(\ref{eq:1}) are, in contrast, unlike any
fundamental interaction that had been probed before the Higgs boson
discovery.
Let us first discuss the Yukawa term.

\subsection{The Higgs Yukawa sector}

Within the SM hypothesis, the Higgs Yukawa term,
$\psi_i y_{ij} \psi_j \phi$, generates masses for all quarks and
charged leptons.
Experimentally, the hypothesis that the Higgs field genuinely produces
these mass terms can be tested by probing $\Hff$ type interactions,
where $f$ is any massive fermion, and verifying the proportionality of
interaction in the amplitude with the fermion mass.
Considering a flavour basis in which the $y_{ij}$ are diagonal, there
are nine independent terms (though one should also check for flavour
changing Higgs interactions, $\Hff'$).

Prior to the discovery of the Higgs boson there was no evidence for
fundamental Yukawa interactions: this was not the part of the SM that
had been probed by 40 years of tests, not even indirectly at LEP.
Discovery provided indirect evidence for two of the nine interactions.
Specifically, the consistency of the cross section in all observed
decay channels was both sensitive to, and consistent with, the SM
expectation for the top and bottom Higgs interactions, given that the
$Htt$ coupling appears in the $gg\to H$ and $H\to \gamma\gamma$
effective interactions, while the $Hbb$ coupling dominates the overall
width of the Higgs boson and so affects all branching fractions and
cross sections.

\begin{figure}[t]
  \centering
  \includegraphics[width=0.32\textwidth]{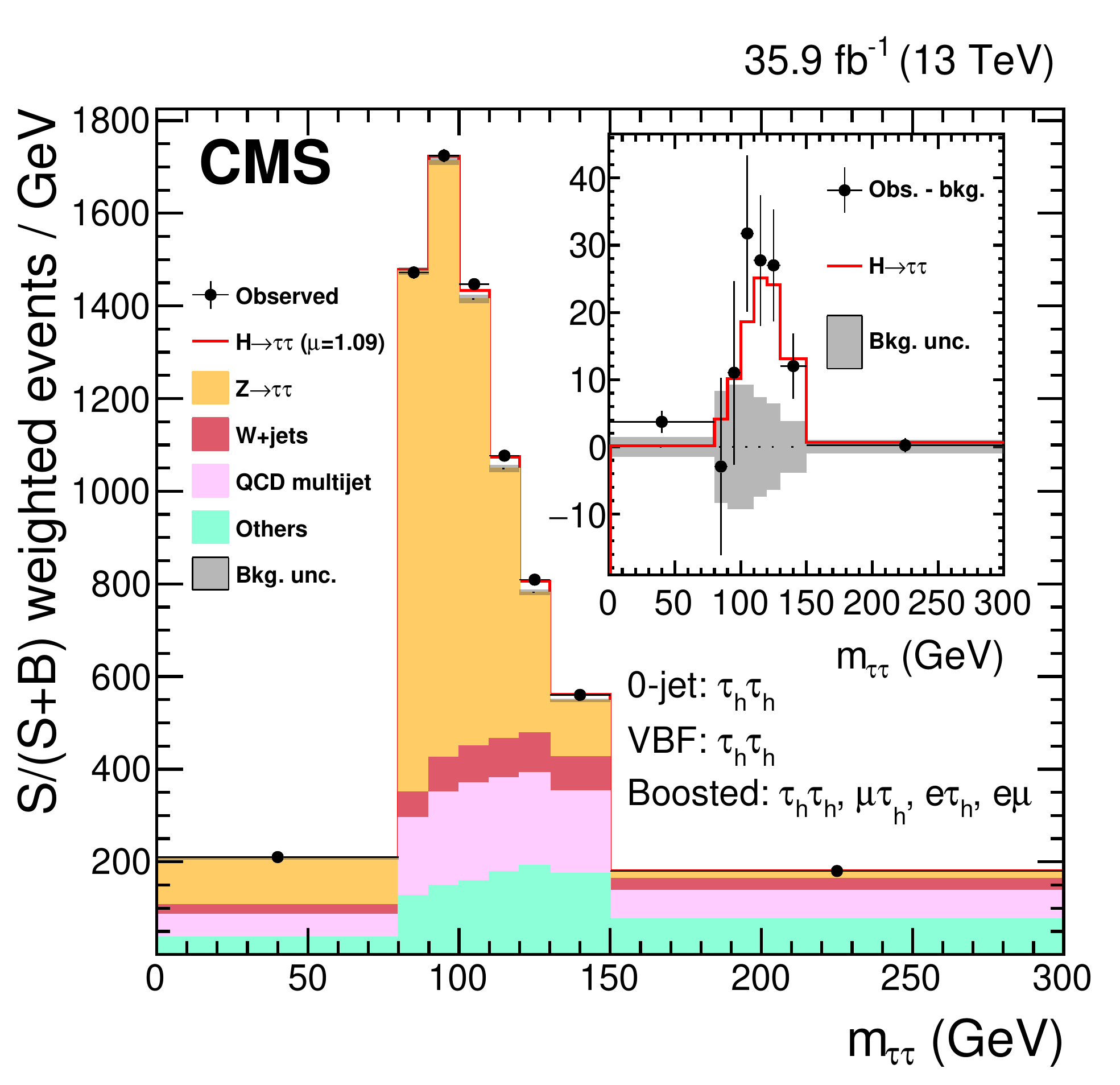}\hfill
  \includegraphics[width=0.32\textwidth,height=0.30\textwidth]{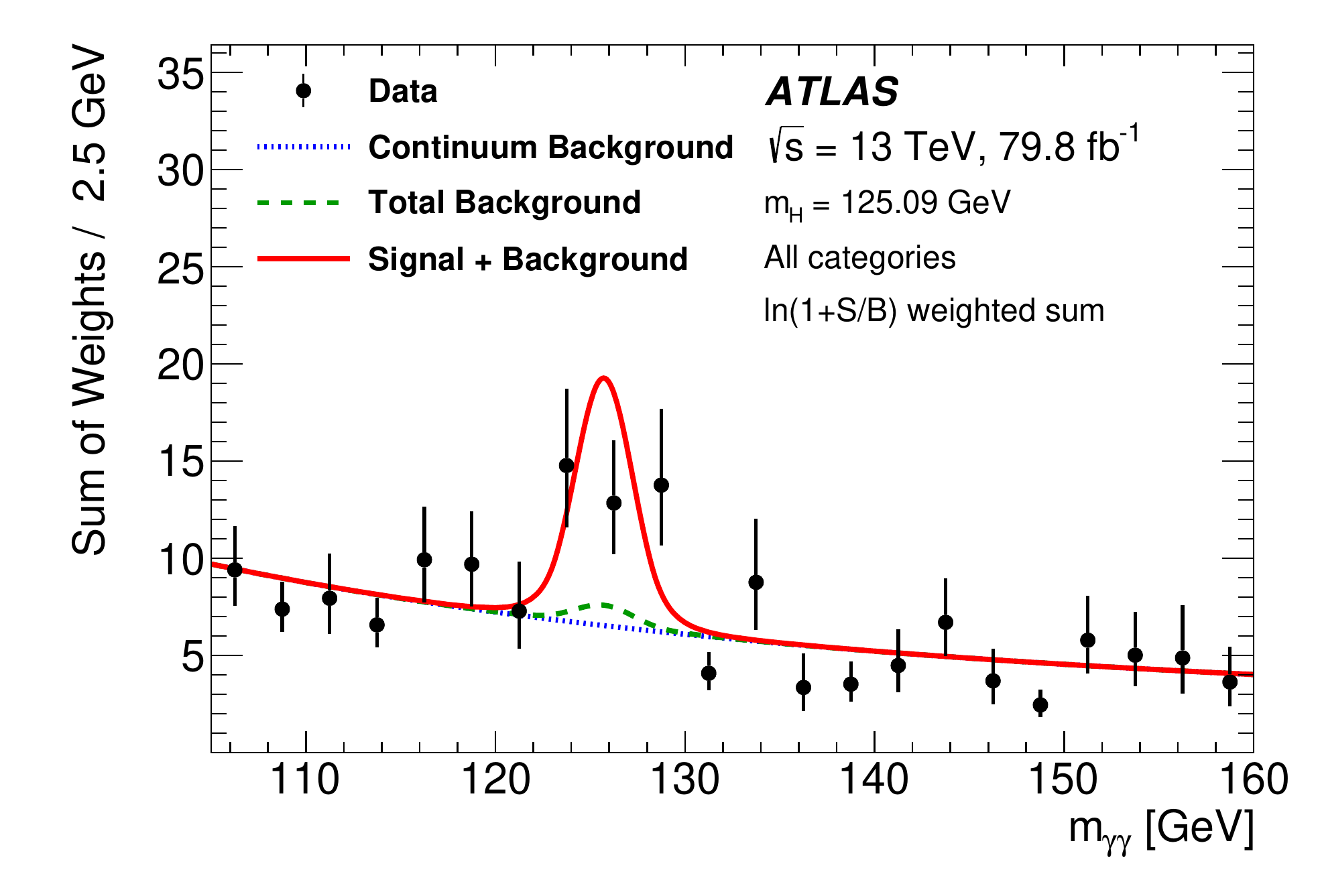}\hfill
  \includegraphics[width=0.32\textwidth]{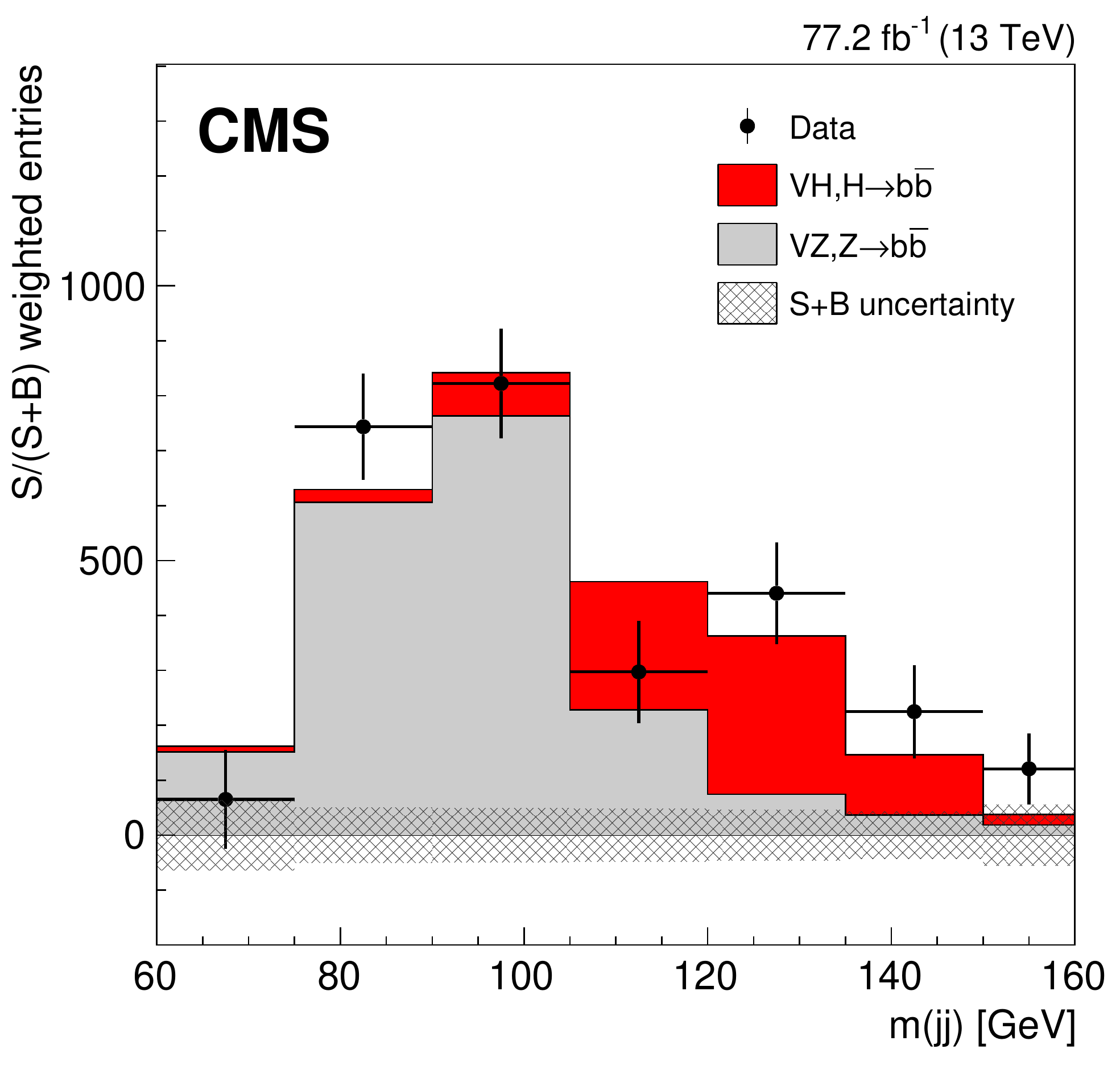}%
  \caption{Selection of plots illustrating the observation of $H \to
    \tau \tau$ (left), the $t\bar t H$ process (middle) and $H \to
    b\bar b$ (right) by the ATLAS~\cite{ATLAS:2018lur,Aaboud:2018urx,Aaboud:2018zhk} and CMS collaborations~\cite{Sirunyan:2017khh,Sirunyan:2018hoz,Sirunyan:2018kst}.}
  \label{fig:higgs-yukawa-discovery}
\end{figure}

Over the past 18 months, our knowledge of Higgs Yukawa interactions
has undergone a revolution, with all three of the (charged-fermion) third-generation
Yukawa couplings now established directly at 5$\sigma$, independently by
each of the ATLAS and CMS collaborations, through the observation Higgs decays to
$\tau^+\tau^-$~\cite{Sirunyan:2017khh,ATLAS:2018lur}, Higgs production
in association with a $t\bar t$
system~\cite{Sirunyan:2018hoz,Aaboud:2018urx} and Higgs decays to
$b\bar b$~\cite{Aaboud:2018zhk,Sirunyan:2018kst}.
A selection of corresponding plots is shown in
Fig.~\ref{fig:higgs-yukawa-discovery}.\footnote{When this talk was
  originally given, only the top and $\tau$ couplings had been
  established.}
This part of the SM is no longer a hypothesis.
It is quite clearly a fact, at least to within the roughly $20\%$
accuracy that accompanies a $5\sigma$ discovery.

What importance should we, as a field, attribute to the observation of
Yukawa interactions?
I would argue that it is comparable to the importance of the discovery
of the Higgs boson in the first place, for three main reasons.

The first reason is that it is exceedingly rare to observe a
qualitatively new kind of interaction.
At the risk of oversimplification, so far matter particles, fermions,
were known to have gravity and gauge interactions; now we know that
they have a third kind, Yukawa scalar interactions.

The second reason is related to the major role that (still
hypothetical) first-generation Yukawa interactions play in the world
we experience every day.
One example concerns the proton--neutron mass difference.
The proton is lighter than the neutron, despite the fact that it is
electrically charged while the neutron is not.
As quantified for example in Ref.~\cite{Borsanyi:2013lga} this is a
consequence of the down quark mass (about $4.7\MeV$) being larger than
the up quark mass (about $2.2\MeV$). Within the SM hypothesis, that
stems from the pattern of first-generation Yukawa couplings.
Were the first generation quark Yukawa couplings to be zero, leading
to correspondingly massless quarks, protons would be unstable because
they would be \emph{heavier} than neutrons.
%
%
As a result there would be no stable hydrogen.\footnote{Deuterium
  would conceivably still be stable.}
The electron's Yukawa coupling is equally important: through its
impact on the electron mass it sets the size of atoms and their energy
levels, the Bohr radius being proportional to $1/m_e$.%
\footnote{A variety of similar thought experiments, specifically about
  a world without a Higgs field, were carried out by Quigg and
  Shrock~\cite{Quigg:2009xr}.
  In contrast to the discussion here, their scenarios have an impact
  also on the electroweak vector bosons.}

A third reason for assigning importance to the observation of Yukawa
couplings is that the mystery of the hierarchy of (charged) fermion
masses is, in the SM hypothesis, converted to a mystery about a
hierarchy of couplings.
Does it matter which of these two we talk about?
My attitude is that it would be good to know, for sure, that we are
indeed dealing with a hierarchy of couplings.
Only then can we be sure of what deeper problem we should be trying to
understand.%
\footnote{To take a simple alternative, the Giudice--Lebedev
  mechanism~\cite{Giudice:2008uua} 
  generates small first and second generation masses by terms in the
  Lagrangian $Y_{ij}(\phi)\,\psi_i \psi_{\!j}\, \phi + \text{h.c.}$
  with
  $Y_{ij}(\phi) = c_{ij}\left(\phi^{\dagger}\phi/M^2\right)^{n_{ij}}$,
  where $M$ is some large new mass scale, which would predict
  enhancement of apparent Higgs-fermion interaction strengths (in the
  amplitude) by a factor $2n_{ij}+1$.
  A more sophisticated realisation of this picture, taking into
  account recent constraints, is discussed in
  Ref.~\cite{Bauer:2017cov}.
  I am grateful to Uli Haisch for discussions on these points.
}

When communicating with our colleagues in neighbouring fields, and
with the general public, we should be wary of talking of vindication of
the standard model and an ensuing dead-end~\cite{NYTimesDeadEnd}.
What we could instead be saying about recent developments is:
\begin{quote}
  The $>5\sigma$ observations of the $ttH$ process and
  $H\to \tau^+\tau^-$ and $H\to b\bar b$ decays, independently by
  ATLAS and CMS, firmly establish the existence a new kind of
  fundamental interaction, Yukawa interactions.
  
  Yukawa interactions are important not merely because they had never
  before been directly observed, but also because they are
  hypothesised to be responsible for the stability of hydrogen, and
  for determining the size of atoms and the energy scales of chemical
  reactions.
  Establishing the pattern of Yukawa couplings across the full
  remaining set of quarks and charged leptons is one of the major
  challenges for particle physics today.
\end{quote}
The second paragraph of that statement serves as a reminder of just
how much more remains to be established before we can truly talk about
the vindication of the Standard Model, even just its Yukawa sector.
In some cases we know how to go about learning more.
For example, the muon Yukawa coupling can be established at the
HL-LHC, through observation of $H\to \mu^+\mu^-$ decays.
Establishing the charm Yukawa coupling probably requires a lepton
collider (see e.g.\ the analysis for the ILC~\cite{Fujii:2017vwa}), or
an $ep$ facility~\cite{Klein:2018rhq}.
Importantly, the total width can also be strongly constrained at
$e^+e^-$ colliders.
For other Yukawa couplings, the path forwards still remains to be
established: ideas for constraining the quark Yukawa sector are
highlighted in the references of Refs.~\cite{Bauer:2017cov,Duarte-Campderros:2018ouv} and it has
even been suggested that the electron Yukawa coupling could be probed
in $s$-channel resonant Higgs production at the
FCC-ee~\cite{dEnterria:2017dac}.

\subsection{The Higgs self coupling}

Aside from the Yukawa interactions, the other qualitatively new
structure in Eq.~(\ref{eq:1}) is 
the Higgs potential term, $V(\phi) = -\mu^2 \phi^2 + \lambda \phi^4$,
which generates $HHH$ interactions.
A theory with $\phi^4$ interactions is familiar insofar as scalar
theory with $\phi^n$ interactions serves for almost every graduate
student's introduction to field theory.
Yet we should not forget that a fundamental $\phi^4$ theory has never
been observed in nature.
Nor is there even any indirect experimental evidence for it.
This part of the SM is not merely not yet knowledge, the non-trivial
part of it, i.e.\ structure beyond the simple existence of a minimum
and associated vacuum expectation value, is totally unexplored.
It is also the keystone of the SM: without the potential, the rest of
the edifice falls down.

\begin{figure}[t]
  \centering
  \includegraphics[width=0.45\textwidth]{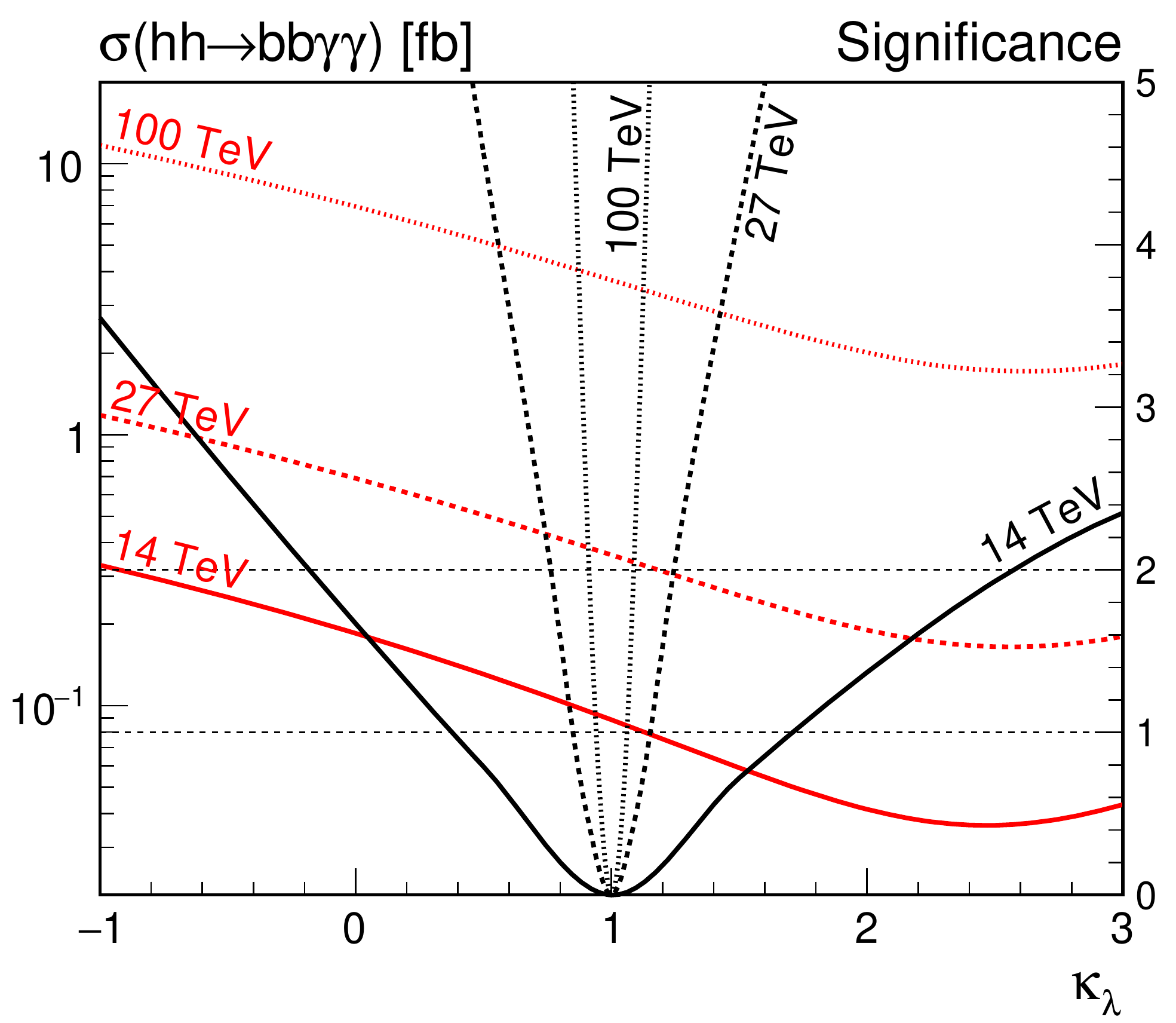}\hfill
  \includegraphics[width=0.5\textwidth]{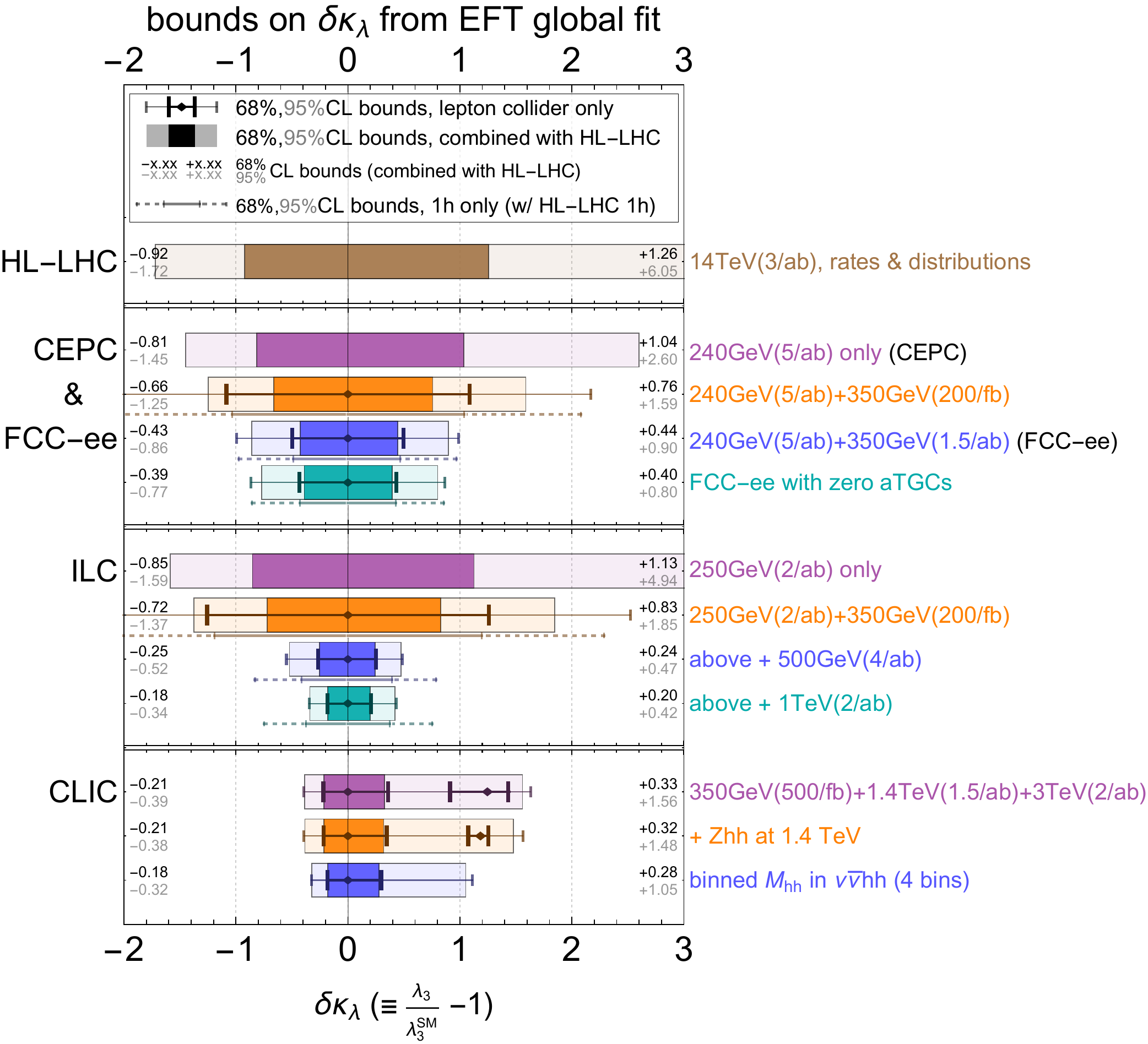}
  \caption{Projected constraints on the coefficient $\lambda$  (or
    rather the factor $\kappa_\lambda$ by which it deviates from the
    SM) of the
    triple-Higgs interaction at a variety of future colliders (this is
    not the same $\lambda$ that appears in Eq.~(\ref{eq:1})).
    Left: at hadron colliders~\cite{Goncalves:2018yva}, showing the
    cross section in red and the contour of constraint significance in
    black (see also Ref.~\cite{Contino:2016spe}).
    Right: at lepton colliders~\cite{DiVita:2017vrr}, compared to the
    HL-LHC projection.}
  \label{fig:self-coupling}
\end{figure}

Together with the Yukawa couplings, determining as much of the
potential as we can should, consequently, be one of the major targets
for the field.
Projected constraints from a range of colliders on the triple-Higgs
interaction strength, which is directly correlated with the structure
of the Higgs potential, are shown in Fig.~\ref{fig:self-coupling}.

\subsection{Higgs outlook}

Overall it seems clear that if we are to push our knowledge of the
Higgs sector to as much of the Lagrangian as is possible, and with
reasonable precision, then in addition to the progress that is
expected to come from the HL-LHC we will need a lepton collider and a
higher-energy hadron collider.

One question whose answer is not yet clear to me, is the precision
that we should seek on our determinations of Higgs interactions, both
the self-coupling and the others.
A threshold of $\sim 20\%$ (on a squared interaction) is akin to the
$5\sigma$ condition for declaring ``observation'' of a new particles,
and seems a minimal requirement.
Once an interaction is established, then should one aim to go the
percent level, to test quantum effects?\footnote{A suggestion made to
  me by M.~McCullough.}
Should one aim to test the coupling over, e.g., an order of magnitude
in transverse momentum, to gain confidence that it is not contaminated
by large higher-dimension operators?
Should models of new physics and the nature of the electroweak phase
transition act as a guide in such circumstances?
Or should we simply set out to measure all fundamental properties of
our universe to some basic precision, say $1\%$, independently of
models for new physics?
These are questions that I believe we as a community need to discuss,
and they feed into the choices that we will argue for as to the next
machine(s) to support.

\section{Searches and anomalies}

\begin{figure}[tb]
  \centering
  \includegraphics[width=0.48\textwidth]{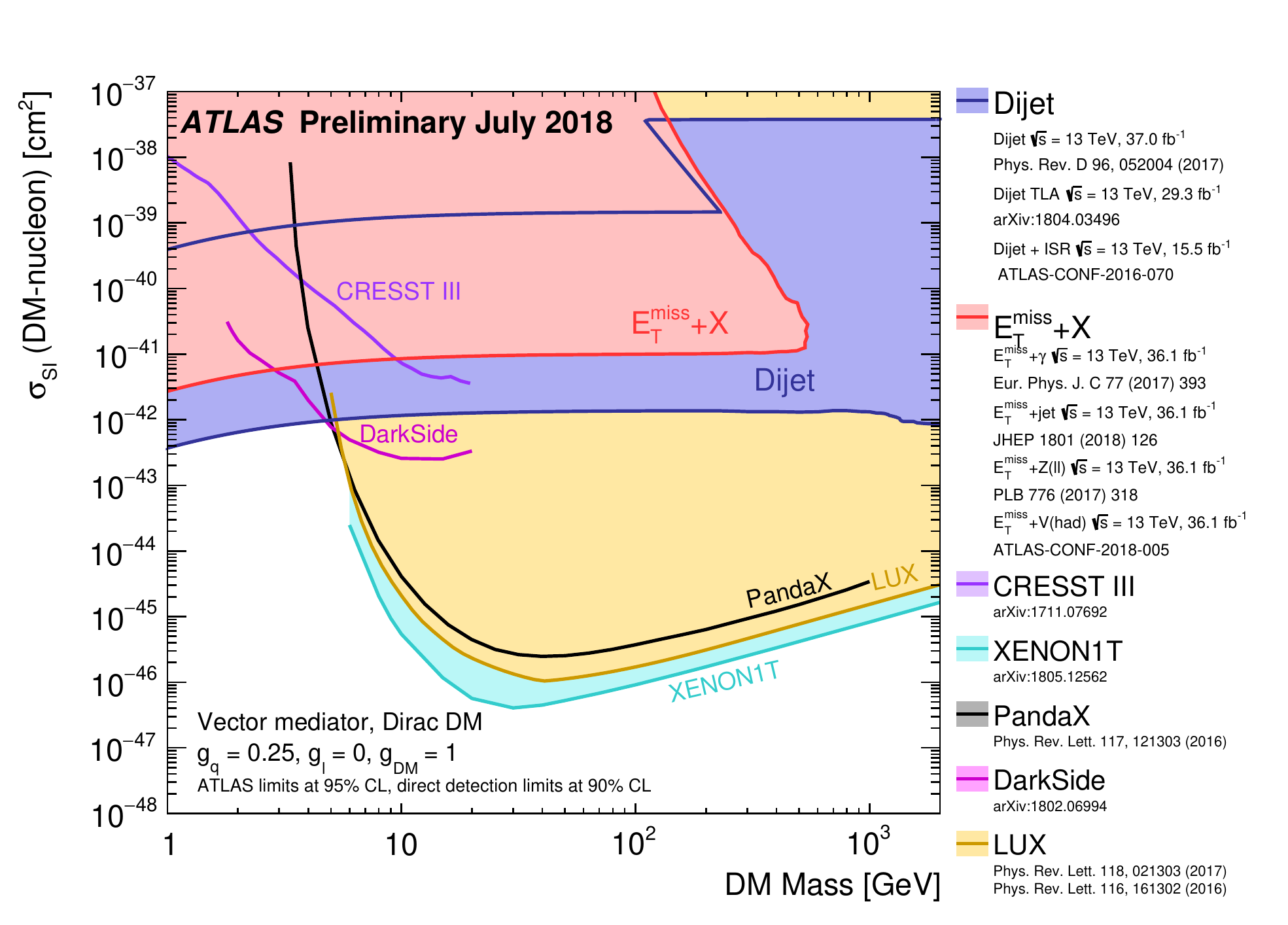}\hfill
  \includegraphics[width=0.48\textwidth]{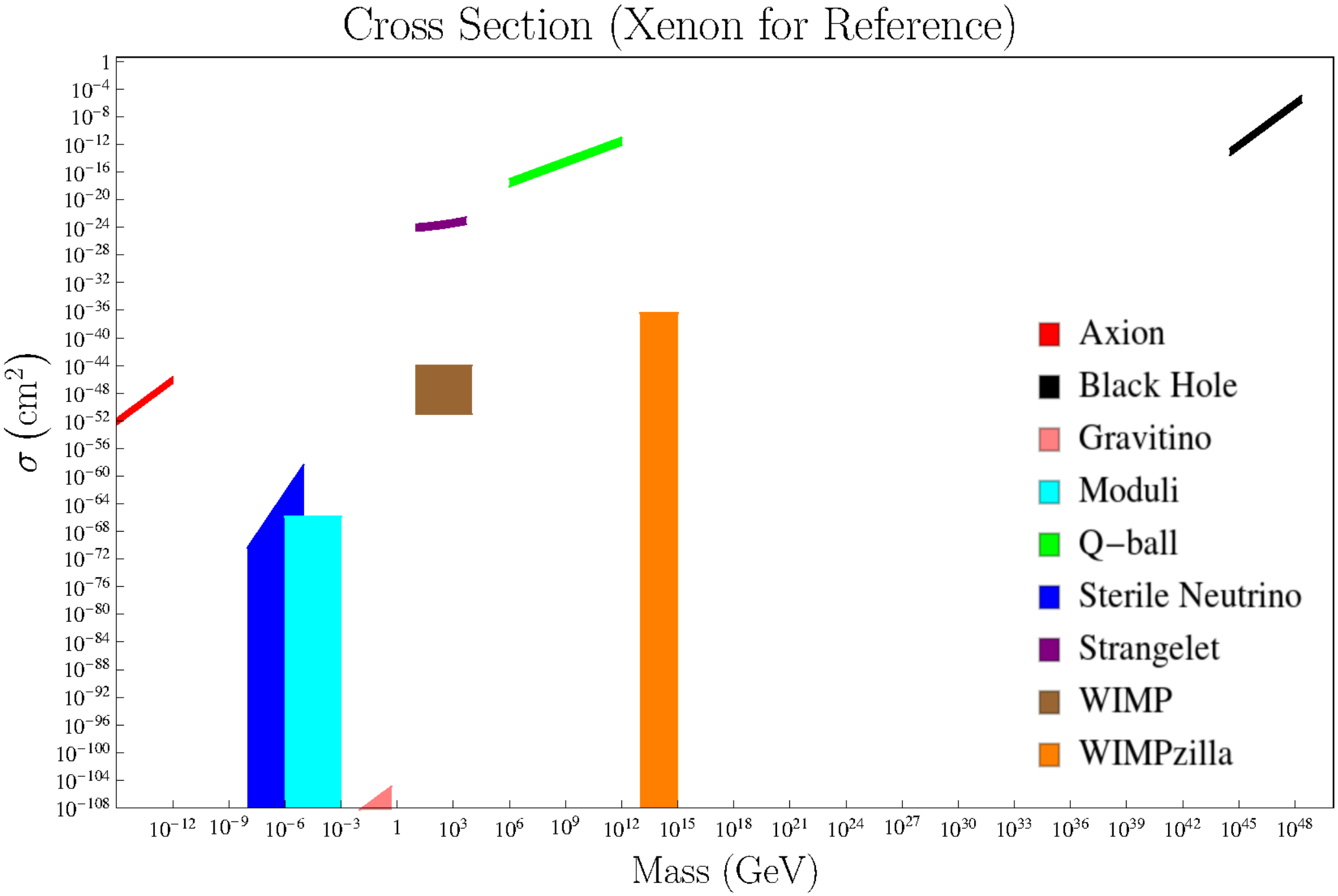}
  \caption{Left: dark-matter exclusion summary plot~\cite{ATLASDM}. The
    comparison between limits from different experiments is model
    dependent. 
    Right: the
    direct-detection cross-section versus mass plane for a set of
    dark-matter models, taken from a Snowmass
    report~\cite{Kusenko:2013saa}.
  }
  \label{fig:dark-matter}
\end{figure}

Prior to the first run of the LHC, many physicists made statements
that they expected the discovery of supersymmetry (e.g.\ Ref.~\cite{Lindau}).
At the start of Run~2, with supersymmetry, and alternatives such as
extra dimensions, proving elusive, some of the emphasis had shifted to
dark matter, especially in communication with the public
(e.g.~\cite{PBS}).
Yet, so far, neither has been found, whether at colliders or
elsewhere, cf.\ e.g.\ the left-hand plot of
Fig.~\ref{fig:dark-matter}.
As already alluded to above, this situation is fostering a quite
widespread sense of dead-end and crisis.

For some time, I have felt that we ought to be more cautious in our
statements of what we expect to find (even if, on occasion, I too have
made statements along similar lines), and of the extent of the crisis
that comes from not having discovered them.
One of the most appealing elements about the fine-tuning argument in
favour of supersymmetry was that such arguments had worked before, for
example in the prediction of the charm quark.
Yet in any prediction by theorists that is not an actual robust proof,
one should allow for the possibility that a new idea could change our
perspective on the classes of solution that might exist for the
underlying problem that they are trying to address, as in some sense
has happened with fine tuning~\cite{Graham:2015cka}.%
\footnote{Incidentally, even had supersymmetry been discovered, it is
  not inconceivable that some people would still have claimed that the
  field is in a crisis. As far back as 1995, statements were being made
  such as ``in the MSSM we have nothing but boring perturbative
  physics to explore below the Planck scale and the interesting
  dynamics of supersymmetry breaking is
  hidden.''~\cite{Nelson:1994ap}.}
As for dark matter, the parameter space and variety of models is vast, as
illustrated in Fig.~\ref{fig:dark-matter} (right).

The point of view that I would advocate is that, as a field, we are trying
to address problems that have a long history of being difficult to solve.
For example, the evidence for dark matter dates back to the 1930's.
We may hold out hope that a given new run of the LHC or of a direct
detection experiment will discover it.
It is natural to be disappointed if that turns out not to happen.
However, we should not be overly surprised if solutions to such
long-standing problems don't fall into our lap each time a new machine
is turned on.
Perhaps, instead, it should serve as a reminder that the immense
challenge of solving such problems is an intrinsic part of the
fascination of the field.

Perhaps another contributing factor to the feeling of crisis comes
from the fact that LHC has now been operating at close to design
energy for a few years and that ``all'' that is left is an increase in
luminosity.
Yet the power of luminosity should not be underestimated, even aside
from the immense progress that it will bring in Higgs-sector studies.
Table~\ref{tab:reach-examples} shows estimates of HL-LHC (exclusion)
sensitivity based on extrapolations from the reach at the time of the
conference.
It also shows what energy collider would be needed to obtain the same
reach by increasing the centre-of-mass energy rather than the
luminosity.
The answer depends significantly on the mass region that is already
being probed.
The chargino example is notable, where the equivalent centre-of mass
energy is over four times the current LHC energy: increased luminosity
buys the field substantial reach.

\begin{table}
  \centering
  \begin{minipage}[c]{0.55\linewidth}
    \begin{tabular}{ccc}\toprule
      LHCP\,2018              &   estimated            & energy needed    \\
      reach                &    HL-LHC reach             & for same reach   \\
      (13 TeV 80 fb$^{-1}$) & (14 TeV,  3ab$^{-1}$)& with 80 fb$^{-1}$\\
      \midrule
      4.7 TeV SSM $Z'$     & 6.7 TeV             & 20 TeV           \\[7pt]
      \shortstack{2 TeV weakly\\[-2pt]coupled $Z'$\vspace{-6pt}}
                           & 3.7 TeV             & 37 TeV           \\[7pt]
      \shortstack{680 GeV\\[-0pt]chargino\vspace{-6pt}}
                           & 1.4 TeV             & 54 TeV           \\[2pt]
      \bottomrule
    \end{tabular}
  \end{minipage}\hfill
  \begin{minipage}[c]{0.4\linewidth}
    \caption{Rough ATLAS/CMS exclusion sensitivities at the time of
      the conference, extrapolation to the HL-LHC, and estimate of the
      corresponding collider centre-of-mass energy that would be
      needed to obtain the same reach if one were to increase the
      collider energy while keeping a fixed luminosity. (HL-LHC and
      high-energy collider estimates made using the collider-reach
      tool~\cite{collReach}.)}
    \label{tab:reach-examples}
  \end{minipage}
\end{table}

Still, the field will have to contend with the fact that over the next
twenty years the rate of logarithmic integrated luminosity increase at
the LHC will be considerably slower than it has been so far.
In such a context, the reduced pressure to update results for each new
conference can perhaps be compensated by further encouraging opportunities for creative
thinking, whether for novel kinds of searches (for example long-lived
particle searches~\cite{LongLived}),
new ways of incorporating machine learning, approaches to automating
searches (e.g.~\cite{Aaboud:2018ufy}), or even finding ways of making
existing LHC data more accessible to researchers who are not
members of a given experiment~\cite{OpenData}.\footnote{Straightforward access to research-usable open
  data clearly remains a challenge for the field, even if some
  progress has been made in the past couple of years by CMS (with the
  data also being used by theorists~\cite{Larkoski:2017bvj}).
  Perhaps one step towards encouraging progress would be for major
  conferences such as LHCP to schedule a dedicated parallel session,
  and associated plenary talk, on developments and issues in making
  research-quality data accessible (the question goes well beyond one of outreach,
  which the session where the topic is currently addressed).
  At the very least, this would help give visibility to those involved
  in this challenging task and communicate the message that the field
  genuinely values the endeavour.
}

This section wouldn't be complete without a mention of a number of
anomalies across different sectors of particle physics: those in the
flavour sector saw much attention at this conference (see Hiller and
Humair's talks~\cite{HillerFlavour,HumairFlavour}).
Others include the longstanding issue in
$g_\mu-2$~\cite{Dorokhov:2016knu}, short baseline neutrino experiment
anomalies, e.g.\ as recently seen by
Miniboone~\cite{Aguilar-Arevalo:2018gpe}, the $16.7\MeV$ Beryllium~8
peak~\cite{Krasznahorkay:2015iga}, the hint of a di-muon peak in
events with a $b\bar b$ in ALEPH and
CMS~\cite{Heister:2016stz,Sirunyan:2018wim}.
One may still be optimistic that one or other of them will, in the
coming years, see the corroborating evidence needed to establish the
existence of laboratory-accessible physics beyond the standard model
(aside from the existence of neutrino masses).

Going forwards, a question that I feel we ought to discuss more is
that of how we prioritise searches in different regions of parameter
space.
At the LHC, this is relatively straightforward: one searches for
everything that can be searched for.
But beyond the LHC, should one evaluate an experiment simply based on
the new area in the $\ln(\text{coupling})$--$\ln(\text{mass})$ plane
that it will cover, multiplied by the number of channels?
Should one prioritise high mass scales, whether on fine-tuning grounds
(there is still a chance that this is the right perspective), or because
most of our discoveries so far have come from exploring ever higher
scales?
%

\section{Precision physics at the LHC}

One of the surprises with the LHC is its remarkable potential for
precision measurements.
These include improved determinations of fundamental parameters such
as the $W$ and top-quark masses, constraints on parton distribution
functions, precise evaluations of backgrounds for missing energy
searches, and more generally constraints on higher-dimensional
operators.
In due course, Higgs physics will also join the precision family.

Precision matters in its own right: the confidence that we have in a
hypothesis that is supported by experimental results to within $1\%$
precision is very different from a situation where we are limited to
$20\%$ tests.
Additionally, precision can contribute significantly to BSM
sensitivity.
As an example, considering constraints on the energy-scale of
dimension-$6$ operators (in processes that involve interference
with SM operators), statistically those constraints scale as the
fourth root of the luminosity,
$\Lambda_\text{dim-6} \sim \mathcal{L}^{\frac14}$.
Going from $80\fb^{-1}$ to $3\ab^{-1}$ could then translate to an
increase in mass scale sensitivity by a factor of almost $2.5$, which
is greater than any of the estimates for direct searches shown in
table~\ref{tab:reach-examples}.

An important question for the future of precision physics at the LHC is our
understanding of, and ability to make progress in reducing systematic
uncertainties, both experimental and theoretical.
As an example, Huss and Melnikov's talks~\cite{Huss,Melnikov} at this
conference summarised the rapid advances that are being made in
calculations of higher-order perturbative QCD and EW contributions to
many LHC processes.
Prestel's talk at this conference~\cite{Prestel} outlined some of the
recent progress in general-purpose Monte Carlo event generators.
This is an important area because differences between generators are
quite often among the dominant systematic in experimental calibrations
and measurements.\footnote{I am perhaps biased here, having recently
  developed an active interest in the topic of what parton shower
  algorithms can and should achieve~\cite{Dasgupta:2018nvj}.
}

Two areas that are potential long-term limiting factors are
non-perturbative corrections on the theoretical side and the
uncertainty in the absolute luminosity on the experimental side (see
for example Valentinetti's talk~\cite{Valentinetti}).
The former has recently seen a pioneering
analysis~\cite{FerrarioRavasio:2018ubr}, sensitive to all main aspects
of the leading (relative $\Lambda_\text{QCD}/m_\text{top}$ suppressed)
non-perturbative contributions in top-quark measurements, both for
cross sections and the reconstructed top mass.
Possible avenues of progress for the latter are discussed in
Ref.~\cite{Dabrowski-HL}. 

\section{Heavy-ion collisions}

Over the past twenty years, heavy-ion collisions have revealed an
array of new and intriguing phenomena.
The latest instalment is the discovery that proton--proton and
proton--lead collisions yield signals of collective effects (elliptic
flow) that had long been thought to be characteristic of larger
systems such as lead--lead.
This was summarised in the talk by Bellini~\cite{Bellini}.
The condition for such signals of collective effects to be observed is
that one should focus on the subset of proton--proton and proton--lead
collisions that have a high final-state multiplicity.
Thus the field of heavy-ion physics is teaching us new ways of looking
at, and new features of, proton--proton collisions.

At the same time, techniques originally developed in the context of
new-physics searches, in particular jet-substructure
techniques~\cite{Larkoski:2017jix} are
today seeing applications in heavy-ion collisions, with the heavy-ion
field being a very fast adopter of the latest developments, as
reviewed in the talk by Apolin\'ario~\cite{Apolinario}. 

One possibility that has seen first feasibility studies recently is
that of using electroweak-scale probes in heavy-ion
collisions~\cite{dEnterria:2017jyt} either to learn more about the
quark--gluon plasma or the electroweak-scale probes themselves.
For example, Ref.~\cite{Berger:2018mtg} has suggested that one may
better measure the coupling of the Higgs to $b$-quarks in heavy-ion
collisions because the medium will quench background $b$-jets and not
those from a (long-lived) Higgs decay to $b\bar b$.
Ref.~\cite{dEnterria:2018bqi} has suggested the Higgs width may be
significantly modified while passing through the medium, though one
should keep in mind that this requires further theoretical analysis,
especially given the subtleties that arise in width calculations at
finite temperature~\cite{Czarnecki:2011mr}.\footnote{I am
  grateful to Urs Wiedemann for pointing out this reference to me. }
I myself have become interested in the question of whether the finite
lifetime of top-quark and $W$ bosons, as well as the finite
decoherence time of the $q\bar q'$ system from $W$ decay could be
exploited to give time-resolved information about the quark--gluon
plasma~\cite{Apolinario:2017sob}.
In all cases, given the small cross sections of electroweak-scale
probes, it seems likely that increased luminosity and/or collider
energy are a necessity for these studies to become reality.
The possibility of obtaining higher nucleon--nucleon luminosities in
collisions of intermediate-size systems~\cite{Jowett} may be important
in this respect.

\section{Conclusions}

Many in our field talk of a period of crisis because the LHC has
``failed'' to discover physics beyond the SM.
I would argue that we might equally think of this as a golden period
of particle physics, in particular from an experimental point of view:
it is not often that physicists gain access to new, apparently
fundamental classes of interaction, and that is precisely what the
discovery of the Higgs boson has made possible for us, in the past 18
months with the observation of the third-generation Yukawa
interactions, in the future with Yukawa couplings beyond the 3rd
generation and the $\phi^4$ structure of the Higgs potential.

Of course, the interactions that are being discovered today are far
from novel from a theoretical point of view, and this is a cause of
despondency for many.
Yet, while it is perfectly legitimate to get excited about
conceptually new realisations of how our universe might be configured,
we should not fall into the trap of failing to be excited when we are
in the process discovering how essential aspects of our universe
actually are configured.
The fact that these aspects were hypothesised long ago does not take
away from their importance.
On the contrary, even if we are still left with big unanswered
questions such as the nature of dark matter and the solution to the
hierarchy problem, we should retain some capacity for wonder and awe
at the fact that what is arguably the minimal hypothesis --- the
Standard Model --- actually seems, for now, to be how the directly
accessible part of nature is arranged.

\section*{Acknowledgements}
I am grateful to Uli Haisch, Matthew McCullough, Michelangelo Mangano,
Guilherme Milhano, Hitoshi Murayama, Urs Wiedemann and Giulia
Zanderighi for stimulating conversations on topics related to this
writeup and/or for helpful comments.
This work is supported in part by a Royal Society Research
Professorship under grant RP{\textbackslash}R1{\textbackslash}180112.


\end{document}